\documentclass[aps,prl,reprint,amsmath,amssymb, showkeys, superscriptaddress]{revtex4-2} 
\usepackage{color,soul} 
\usepackage{mathtools}
\usepackage{float}
\usepackage{graphicx}
\usepackage{amsmath}
\usepackage{amsthm}
\usepackage{amssymb}
\usepackage{multirow} 
\usepackage[normalem]{ulem}
\usepackage{lineno}
\usepackage{chemformula}
\usepackage[T1]{fontenc} 
\begin{document}

\title{Entropy exchange in an inter-correlating binary quasi-classical system: Concept of entropy-bath accelerated molecular dynamics}
\author{Projesh Kumar Roy}
\affiliation{Department of Chemistry, National Institute of Technology Tiruchirappalli, Tanjore Main Road, NH83, Tiruchirappalli, Tamil Nadu 620015, India}
\email{projesh@nitt.edu}

\date{\today}


\begin{abstract}
This letter highlights the entropy exchange phenomenon in a coupled binary inter-correlating system following Haldane's non-linear statistical correlation. A unique coupling between a classical and a quantum-like system at the marginal distribution is observed. It is shown that the quantum nature of a system can arise without any self-correlation. Extending this idea, an enhanced sampling method in molecular dynamics simulation is postulated where a classical system is forced to show quantum-like behavior with the help of an {\it entropy-bath}. An entropy-bath exchanges entropy with the system to scale the potential energy distribution of the system, so that a probability upper bound at each energy level is maintained. An algorithm to implement the entropy-bath accelerated molecular dynamics simulation is discussed. Using low temperature vitreous silica as an example, the capability of such an algorithm to greatly improve sampling of the potential energy landscape under equilibrium conditions for kinetically arrested systems is highlighted.  
\end{abstract}
 
\maketitle


\section{Introduction}

Intermediate statistics can be described as a unified statistical theory that encompasses Fermi-Dirac (FD), Bose-Einstein (BE), and Maxwell-Boltzmann (MB) statistics. Real-world applications of intermediate statistics have been actively explored in recent times. Among several variants of intermediate statistics~\cite{Niven_EurPhysJB_2009, Niven_Grendar_PhysLettA_2009, Ourabah_Tribeche_PhysRevE_2018, Abutaleb_IntJTheoPhys_2014, Yan_PhysRevE_2021}, Haldane-Wu (HW) statistics~\cite{Haldane_PhysRevLett_1991, Wu_PhysRevLett_1994}, and its updated version Polychonakos statistics~\cite{Polychronakos_PhysLettB_1996, Polychronakos_NuclPhysB_1996}, have shown promising applications in quantum~\cite{Veigy_Ovury_ModPhysLettA_1995_1, Veigy_Ovury_ModPhysLettA_1995_2, Camino_Goldman_PhysRevB_2005, Arovas_Wilczek_PhysRevLett_1984} as well as quasi-classical systems~\cite{Cerofolini_JPhysA_2006, Davila_Pastor_JChemPhys_2009, Fernandez_Pastor_ChemPhysLett_2014, Riccardo_Pasinetti_PhysRevLett_2019, Mezzasalma_RSCAdv_2019, Sharma_Muller_JStatMech_2015}. The concept of HW statistics is based on the idea of Haldane's statistical correlation theory~\cite{Haldane_PhysRevLett_1991} in quantum systems, which was extended and applied to BE statistics. Recently, I have shown~\cite{Roy_arxiv_2024} that a non-linear extension of Haldane's theory can be applied to classical MB statistics to derive a novel quasi-classical intermediate statistics for a pure self-correlating system. It has a similar form as the classical fractional exclusion statistics~\cite{Roy_PhysRevE_2022} (CFES), which was derived earlier for a classical system without any quantum approximation. 

In this letter, I show that self-correlation is not necessary to show quantum-like characteristics in this quasi-classical model. Using a binary inter-correlating system, I prove that due to inter-correlation, one part of the system can show classical distribution, whereas the other part can show quantum-like exclusion characteristics. This unique coupling paves the way to conceptualize an {\it entropy-bath} in molecular dynamics (MD) simulation, where a classical system is forced to behave as a quantum-like system obeying the exclusion principle by exchanging entropy with a hypothetical reservoir. The letter is organized as follows: In section II~\ref{sect:theory}, I derive the coupled energy distributions for an inter-correlated binary system, and I explain the entropy exchange between the coupled systems. Following this, I propose a possible application of the binary system in MD simulation as an entropy-bath for a faster exploration of the potential energy landscape (PEL). In section III~\ref{sect:results}, I show the effects of entropy-bath accelerated MD (EBMD) in glassy Silica. Finally, I conclude in section IV~\ref{sect:conclusion} with an outlook. 


\section{Theory and Derivation}
\label{sect:theory}

\textit{Inter-Correlating Binary System}\label{subsect:binary_corr}: In all the following sections of this letter, the word \textit{particle} will be used to define the building blocks of a system, whose energy distribution follows the maximum entropy (MaxEnt) principle. I consider a closed system consisting of two species; $Q$ and $S$; where neither component has any self-correlation, but they are inter-correlated to each other. Particles belonging to both $Q$ and $S$ components are distinguishable objects and follow MB statistics, i.e., their probability distributions ($p$) follow $p= g\exp\{-\mu-\beta\epsilon\}$; where $\beta=(1/k_BT)$ with $k_B=$ Boltzmann constant and $T=$ temperature of the system, $\epsilon$ is the energy of the system with degeneracy $g$, and $\mu$ is a constant related to chemical potential in thermal ensembles. The components contain a total of $N=N_Q+N_S$ particles with $N\mu=N_S\mu_S+N_Q\mu_Q$, and total energy $E=N_QE_Q+N_SE_S$, respectively. Let's imagine that, at the marginal distribution, the population $n_q$ at energy level $\epsilon_q$ of component $Q$ affects the population $n_s$ at energy level $\epsilon_s$ of component $S$. I describe the nature of this correlation using Haldane\rq{s} non-linear statistical correlation theory~\cite{Roy_arxiv_2024},

\begin{equation}
\tilde{g}_s = g_s - \sum_q \gamma_{sq}\Bigg(\frac{n_q^m}{g_q^{m-1}}\Bigg)
\label{eqn:nonlinear_haldane_coupled} 
\end{equation}

$\tilde{g}_s$ and $g_s$ are the available and true degenerate sates of $S$ component at energy level $\epsilon_s$, respectively. $m$ is an integer~\cite{Roy_arxiv_2024} with range $1  \leq m \leq \infty$. The correlation function, $\gamma_{sq}$, can be a function of the properties of both $S$ and $Q$ components. The total number of microstates ($W$) for such a coupled binary quasi-classical system can be written as,

\begin{equation}
W = W^{\text{MB}}_SW^{\text{MB}}_Q\prod_s \Bigg\{1-\frac{1}{g_s}\sum_q\gamma_{sq}\Bigg(\frac{\nu_q^m}{g_q^{m-1}}\Bigg)\Bigg\}^{\nu_s}
\label{eqn:microstates}
\end{equation}

$\nu_s$ and $\nu_q$ are un-optimized populations and $W^{\text{MB}}_S= g_s^{\nu_s}/\nu_s!$, $W^{\text{MB}}_Q= g_q^{\nu_q}/\nu_q!$ for $S$ and $Q$ components, respectively. Next, the MaxEnt method is used to obtain the optimal distribution of the combined system under NVE conditions, where the $E$ and $\mu$ are constant. Using the MaxEnt method, we can obtain the partial derivatives of $W$ of this system, $(\partial \ln W/\partial \nu_s)$ and $(\partial \ln W/\partial \nu_q)$ to maximize $W$ and obtain the optimal populations for the $S$ and $Q$ components for the combined system as $\nu_{q,\text{max}} \equiv n_q$ and $\nu_{s,\text{max}} \equiv n_s$, which can be mathematically expressed as,

\begin{equation}
n_s = g_sA_s\Bigg\{1 - \frac{1}{g_s}\sum_q\gamma_{sq}\Bigg(\frac{n_q^m}{g_q^{m-1}}\Bigg)\Bigg\}
\label{eqn:S_distribution_orig}
\end{equation}

\begin{equation}
n_q = g_qA_q\exp\Bigg\{-m\Bigg(\frac{n_q^{m-1}}{g_q^{m-1}}\Bigg)\sum_s \Bigg(\frac{\gamma_{sq}n_s}{\tilde{g}_s}\Bigg)\Bigg\}
\label{eqn:Q_distribution_orig}
\end{equation}

where $A_s = \exp(N_S\mu_S-\beta\epsilon_s)$, $A_q = \exp(N_Q\mu_Q-\beta\epsilon_q)$. To separate $s$- and $q$-dependent terms in the summand in Eq.~\ref{eqn:Q_distribution_orig}, I make an approximation that $\gamma_{sq}$ is separable as, $\gamma_{sq} \equiv \gamma_s\gamma_q$. For dilute systems; i.e., $g_s >> n_s$ and $g_q >> n_q$; the final equation for the energy distribution of the $Q$ component reads,

\begin{equation}
n_q = g_qA_q\Bigg\{1 - m\langle\zeta_S\rangle\gamma_q\Bigg(\frac{n_q^{m-1}}{g_q^{m-1}}\Bigg)\Bigg\}
\label{eqn:Q_distribution_final}
\end{equation}

$\langle\zeta_S\rangle$ is effectively an average of the $\gamma_s/\tilde{g}_s$ factor. At the marginal distribution where the entropy of both components is maximized, I can substitute the values of Eq.~\ref{eqn:S_distribution_orig} in Eq.~\ref{eqn:Q_distribution_orig}. Thus, for classical systems with continuous energy levels, $\langle\zeta_S\rangle$ can be written as,

\begin{equation}
\langle\zeta_S\rangle(\mu_S, \beta) = \sum_s\frac{n_s\gamma_s}{\tilde{g}_s} \approx \int_0^{\infty} \gamma_sA_s \partial \epsilon_s
\label{eqn:zeta_S}
\end{equation}

Thus, $\langle\zeta_S\rangle$ is a function of the temperature and $\mu$ of the $S$ component only and therefore a constant for thermal ensembles. $\langle\zeta_S\rangle$ must have a real and finite value for a realistic $n_q$ distribution. It follows that $\gamma_s\exp(-\beta\epsilon_s)$ must converge at $\epsilon_s \to \infty$ following Eq.~\ref{eqn:zeta_S}. 

Interestingly, Eq.~\ref{eqn:Q_distribution_final} takes the form of a CFES-type equation~\cite{Roy_PhysRevE_2022, Roy_arxiv_2024} for $m \geq 2$, where $\langle\zeta_S\rangle$ can be considered as the third Lagrange parameter and $m$ being the exponent as described in reference~\citenum{Roy_PhysRevE_2022}. As a result, there is an upper bound ($n^*_q$) to the particle distribution at each degenerate state $g_q$ for $\gamma_s\gamma_q > 0$ as,

\begin{equation}
n^*_q = \Bigg(\frac{1}{m\langle\zeta_S\rangle\gamma_q}\Bigg)^{\Big(\frac{1}{m-1}\Big)}
\label{eqn:upper_bound}
\end{equation}

Unlike traditional quantum systems, $n^*_q$ is not a universal constant and may be a function of thermodynamic variables such as temperature and pressure due to the presence of $\langle\zeta_S\rangle$ in Eq.~\ref{eqn:upper_bound}. In retrospect~\cite{Roy_PhysRevE_2022, Roy_arxiv_2024}, as $\langle\zeta_S\rangle$ is constant at the optimal distribution, the $n_q$-distribution has an overall constraint $C_Q$ as,

\begin{equation}
C_Q =\sum_q\gamma_q\Bigg(\frac{n_q^m}{g_q^{m-1}}\Bigg)
\label{eqn:Q_constraint}
\end{equation}

Interestingly, the exponent in Eq.~\ref{eqn:Q_distribution_final} and~\ref{eqn:Q_constraint} is same as the purely classical CFES system described in reference~\citenum{Roy_PhysRevE_2022}, and not shifted by 1 as was the case for the self-correlating quasi-classical system described in reference~\citenum{Roy_arxiv_2024}. Substituting Eq.~\ref{eqn:Q_constraint} into Eq.~\ref{eqn:S_distribution_orig}, I get,

\begin{equation}
n_s = g_s A_s\Bigg\{1 -\Bigg(\frac{\gamma_s C_Q}{g_s}\Bigg)\Bigg\}
\label{eqn:S_distribution_final}
\end{equation}

For the linear correlation case at $m=1$, Eq.~\ref{eqn:Q_distribution_final} predicts a temperature-dependent correction on the degeneracy factor of the $Q$ component. However, for non-linear correlation cases; i.e., $m \geq 2$; $n_q$-distribution starts to show quantum-like behavior with $n_q^* > 0$. Interestingly, although the degeneracy factor of the $S$ component is influenced by the $Q$ component (not the other way around), the $n_s$-distribution does not show any quantum-like behavior at all values of $m$ in Eq.~\ref{eqn:S_distribution_final}, as $C_Q$ is a constant for constant $\langle\zeta_S\rangle$. Therefore, this inter-correlating binary quasi-classical system shows a unique coupling between a classical and a quantum-like component in the absence of self-correlation.


\textit{Entropy Exchange}\label{subsect:entropy_exchange}: The Gibbs equation of the total chemical potential for the inter-correlated system can be written as, $N\mu = E + N/\beta - S/\beta$. From Eq.~\ref{eqn:microstates} the total entropy ($S$) can be written as $S = \ln(W_S^{\text{MB}}) + \ln(W_Q^{\text{MB}}) + \chi$, where $\chi$ can be called the {\it exclusion entropy}~\cite{Roy_PhysRevE_2022}, and it has the form,

\begin{equation}
\chi = \sum_s n_s\ln\Bigg\{1-\Bigg(\frac{\gamma_sC_Q}{g_s}\Bigg)\Bigg\}
\label{eqn:exclusion_entropy}
\end{equation}

It is not possible to separate the contributions to $\chi$ into $S$ and $Q$ components from Eq.~\ref{eqn:exclusion_entropy}. This makes the entropy of the total system non-additive, which is a signature for correlated systems. The correlation between $S$ and $Q$ components can be better understood using an equilibrium effective temperature approximation, originally introduced for 2D-silica systems in reference~\citenum{Roy_Heuer_PhysRevLett_2019, Roy_Heuer_JPhysCondMat_2019} and for CFES-type systems in reference~\citenum{Roy_PhysRevE_2022}. For dilute systems, exclusion entropy can be approximated as,                                                                                                                                                                                                                                                                                                                                                                                                                                                                                                 

\begin{equation}
\frac{\chi}{\beta} = \lambda_{SQ}(m,\beta)E_QE_S
\label{eqn:effective_temperature}
\end{equation}

$\lambda(m,\beta)$ is a property of both $S$ and $Q$ components. Substituting Eq.~\ref{eqn:effective_temperature} into the free energy equation, the total energy of the system can be written as $E=E_S+E_Q(1-\lambda_{SQ}E_S)$. Therefore, the $n_q$-distribution can be approximated as, 

\begin{equation}
n_q \propto g_qe^{-\beta(1-\lambda_{SQ}E_S)\epsilon_q}
\label{eqn:effective_temperature_Q}
\end{equation}

$\beta(1-\lambda_{SQ}E_S)$ is the effective inverse temperature factor of the system. As the $Q$ component is correlating with the $S$ component, I have attached $\lambda_{SQ}$ part with $E_Q$ instead of $E_S$. Eq.~\ref{eqn:effective_temperature_Q} describes how $Q$ and $S$ components are correlated to each other. From Eq.~\ref{eqn:exclusion_entropy}, I can approximate the sign of $\lambda_{SQ}$ as $\lambda_{SQ}/|\lambda_{SQ}| \sim -\langle\zeta_S\rangle C_Q/|\langle\zeta_S\rangle C_Q|$. If $\lambda_{SQ} < 0$, any instantaneous increase in $E_S$ will result in a decrease in the effective temperature in $n_q$-distribution, which in turn, will decrease overall energy $E_Q$; i.e., they will be anti-correlated. However, if $(1/E_S) > \lambda_{SQ} > 0$, an increase in $E_S$ will also increase the effective temperature of the $Q$-component and increase $E_Q$; i.e. they will be positively correlated. Hence, any energy fluctuation in $E_S$ will directly affect $E_Q$. 

Thus, the correlation between the coupled systems forces them to continuously exchange entropy in the form of effective temperature to maximize the system entropy in equilibrium conditions. If the size of the $S$-component is sufficiently large, one can assume that any fluctuation of $E_S$ is negligible as compared to $E_Q$. In this special case, one can study the $Q$ component approximately independent of the $S$ component. This concept gives rise to the idea of an {\it entropy-bath} similar to the thermal baths often used in MD simulations, which I elaborate on the next section. 


\begin{figure}[h]
\centering
\includegraphics[scale=0.26]{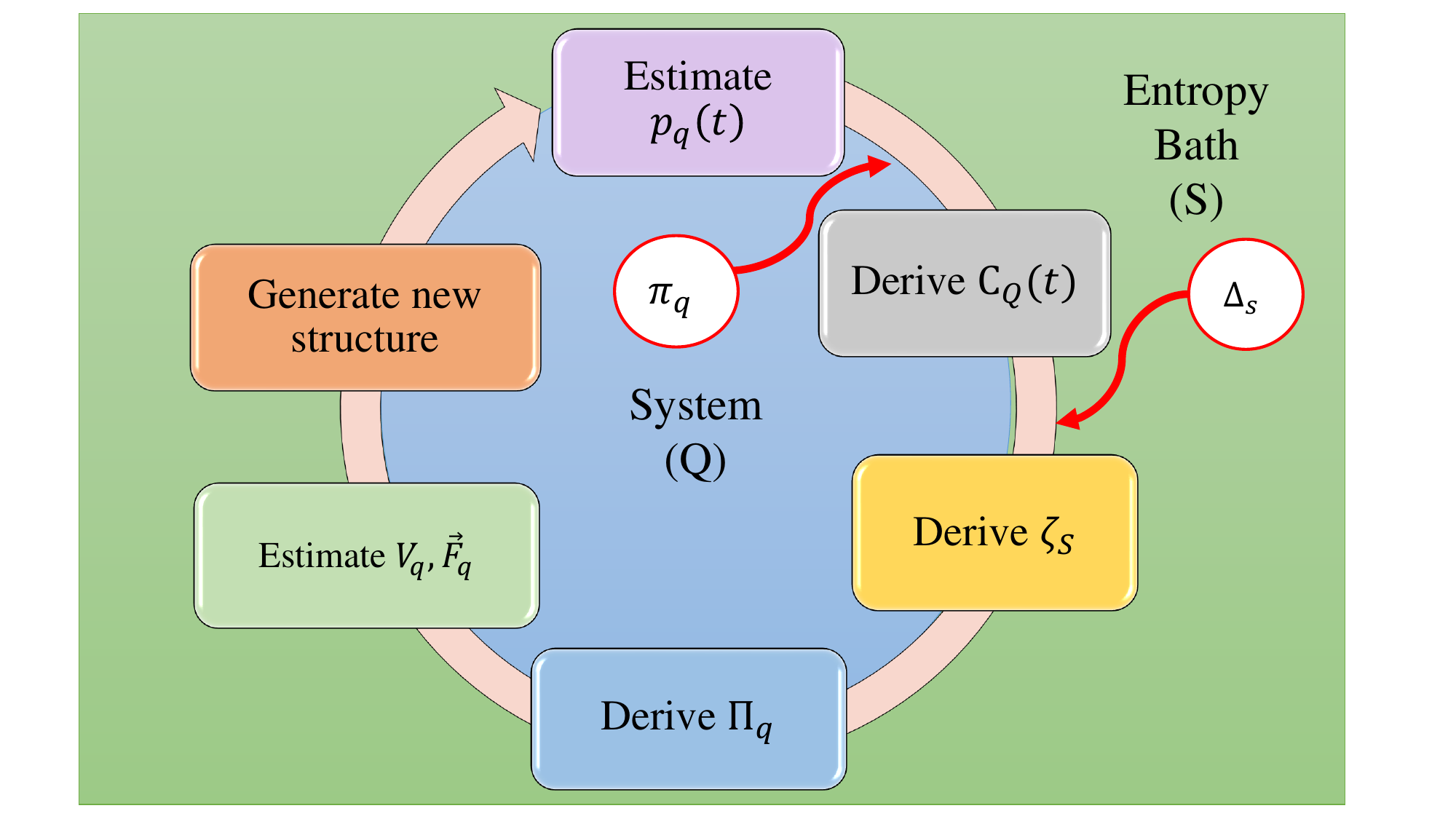}
\caption{General scheme for implementing an entropy-bath in a molecular dynamics simulation.}
\label{fig:algorithm}
\end{figure}

\textit{Concept of entropy-bath}\label{subsect:entropy_bath}: Equilibrium sampling from MD simulation is often hindered by the overpopulation problem. For example, in glassy systems, the potential energy barriers at the low energy range are quite large~\cite{Grigera_Parisi_PhysRevLett_2002} and are often inaccessible by thermal fluctuations. This results in an over-population at a certain minimum, which fails to equilibrate the system within an appreciable time window. Enhanced sampling techniques, such as Accelerated MD~\cite{Hamelberg_McCammon_JChemPhys_2004}, Parallel Temperaring~\cite{Marinari_Parisi_EurPhysLett_1992, Swendsen_Wang_PhysRevLett_1986}, Metadynamics~\cite{Bussi_Parrinello_PhysRevLett_2006, Leone_Parrinello_CurrOPStrBiol_2010}, Replica-Exchange~\cite{Sugita_Okamoto_ChemPhysLett_1999}, Umbrella Sampling~\cite{Torrie_Valleau_JCompPhys_1977, Shankar_Kollman_JCompChem_1992}, Self-Guided MD~\cite{ Wu_Wang_JChemPhys_1999, Wu_Brooks_JChemPhys_2020}, etc. usually require multiple walkers and intensive computational resources and time. Recently, the OneOPES~\cite{Invernizzi_Parrinello_JChemTheoComp_2022, Valerio_Luigi_JChemTheoChomp_2023} method is formulated where the activation energy is directly derived from on-the-fly probability distribution of the system, to quickly converge to a target probability distribution.

The concept of entropy-bath originates from the thought: If a classical system is forced to obey Fermionic quasi-classical distribution at equilibrium (i.e. $\lambda_{SQ} < 0$ and $n_q^* > 0$), its occupancy at a certain potential energy level will be limited. This overpopulation problem will be encountered by the entropy-bath, which will provide activation energy to the system to help it shift to the next energy level. This way, one can increase the sampling range of the PEL to a great extent by using only a single walker. The activation can be provided to the whole dynamics of the system or through specific collective variables (CVs) similar to other enhanced sampling methods like metadynamics. However, the population at such CVs should follow $n_q \to 0$ at $f_{CV} \to \infty$. Another advantage of using an entropy-bath is that a simple re-weighting scheme is needed to extract the density of states (DOS) of the system. 

In MD simulations of thermal ensembles, the system's kinetic energy is controlled by a temperature bath to maintain a constant average value of the second Lagrange parameter, $\beta$. Similarly, an entropy-bath can be thought to control the system's potential energy distribution to maintain a constant average value of the third Lagrange parameter, $\langle \zeta_S \rangle$ at a constant $\beta$. In the present example, we assume a case of thermal ensemble where both $S$ and $Q$ components are in thermal equilibrium. Here, the $S$ component--assumed to be large with access to enormous entropy--may act as an entropy-bath to the $Q$ component, which will scale the potential energy distribution of the $Q$ component during the time evolution of the system with the help of activation energy. If $m\gamma_q/g_q^{(m-1)}$ factor is known beforehand and the equilibrium is achieved, one can directly use Eq.~\ref{eqn:Q_distribution_final} to extract $g_q$ ($\equiv$ DOS for thermal ensembles) of the system. In the following paragraphs, I describe a general scheme to set up the entropy-bath for a thermal ensemble, where the total potential energy of the system, $E_q$, fluctuates and is coupled to the entropy-bath for activation.


\textit{EBMD Algorithm}\label{subsect:algorithm}: For a complex molecular system evolving under a thermal ensemble, the DOS distribution ($G_q (E_q)$) is usually not known \textit{a priory}. To circumvent this issue, Eq.~\ref{eqn:Q_distribution_final} is re-written as, $n_q \propto G_q(E_q)\exp\{-\beta(E_q+V_q)\}$, where $V_q$ is the activation energy provided by the entropy-bath to maintain the exclusion mechanism. It can be expressed as,

\begin{equation}
V_q = -\frac{1}{\beta} \ln\Bigg\{1 - m\langle\zeta_S\rangle\Bigg(\frac{p_q}{\pi_q}\Bigg)^{(m-1)}\Bigg\}
\label{eqn:activation_energy}
\end{equation}

Substituting the probabilities $p_q=n_q/M_Q$ in place of $n_q$ in Eq.~\ref{eqn:activation_energy} makes the energy distribution independent of the total simulation time at equilibrium. Here, $M_Q$ defines the total number of visits at different energy levels of the PEL of $Q$ component, which is proportional to the simulation time. I define two variables $\pi_q(E_q)$ and $\Pi_q(E_q)$ ({\it aka} reference probability) as, 

\begin{equation}
\pi_q(E_q) = \frac{G_q(E_q)}{M_Q\gamma_q(E_q)^{(1/(m-1))}} 
\label{eqn:pi_q}
\end{equation}

and
 
\begin{equation}
\Pi_q(E_q) = \frac{\pi_q(E_q)}{(m\langle\zeta_S\rangle)^{(1/(m-1))}}
\label{eqn:Pi_q}
\end{equation}

for easier referencing. $\pi_q, \Pi_q$ are properties specific to the $Q$ component (Note: Although they are termed as ``probability'', they are actually arbitrary constants). For $m=1$, $V_q$ becomes a constant of energy and thus doesn't contribute to the force calculations. However, for $m > 1$, using a proper choice of the function $\pi_q(E_q)$, one can set a sufficiently large activation energy as $(p_q(E_q)/\Pi_q(E_q)) \to 1$. This excess energy will force the system to cross over to a nearby minimum in the PEL. 

Here, one has to consider that $\pi_q(E_q)$ should not be a dynamic quantity if $\langle \zeta_S \rangle$ is kept constant. If $\pi_q(E_q)$ is constant with time, it follows that $M_Q$ should also be a constant. However, $M_Q$ will always increase with time, and at the thermodynamic limit (i.e., large time) $M_Q$ should be infinity; leading to inconsistencies in the simulation. Therefore, the EBMD algorithm should have manual control over $M_Q$ value so that it doesn't diverge to infinity with time while keeping $p_q$ distribution close to the thermodynamic average. Thus, $\pi_q$ and $M_Q$ cannot truly be constant in EBMD simulation. 

Once $\pi_q(E_q)$ factor is set, one can derive $C_Q = M_Q\sum_q p_q(E_q)(p_q(E_q)/\pi_q(E_q))^{(m-1)}$ and $\langle\zeta_S\rangle$ as,

\begin{equation}
\langle\zeta_S\rangle = \sum_s \frac{p_s}{\Delta_s - C_Q/M_S}
\label{eqn:zeta_S_new}                                                                                                                                                                                                
\end{equation}

respectively, at any point in time $t$. During the simulation, the total energy of the entropy-bath and the system will be sampled simultaneously. Hence, we can set $M_Q=M_S$. $\Delta_s = g_s/(\gamma_sM_S)$ is a property of the entropy-bath. The value of $\Delta_s$ will be estimated at the beginning of the simulation from Eq.~\ref{eqn:zeta_S_new}. The probability distribution $p_q(E_q)$ and $\langle\zeta_S\rangle$ will be updated at every step of the MD simulation. Each particle of the system will receive an additional force $\vec{F}_q = -(1/N_Q)(dV_q/d\vec{r})$, and the particle position and momentum will be updated. After a bath-resetting time $\tau$, $\Delta_s$ will be reset to replace the current $\langle\zeta_S\rangle$ with the original value following Eq.~\ref{eqn:zeta_S_new}. This cycle will continue till a constant value of $\langle\zeta_S\rangle$ is reached and $(p_q(t)/\Pi_q(t)) < 1$ throughout the simulation; which will ensure equilibrium. 


\section{Results and Discussion}
\label{sect:results}

\begin{figure*}[!htb]
\centering
\includegraphics[scale=0.21]{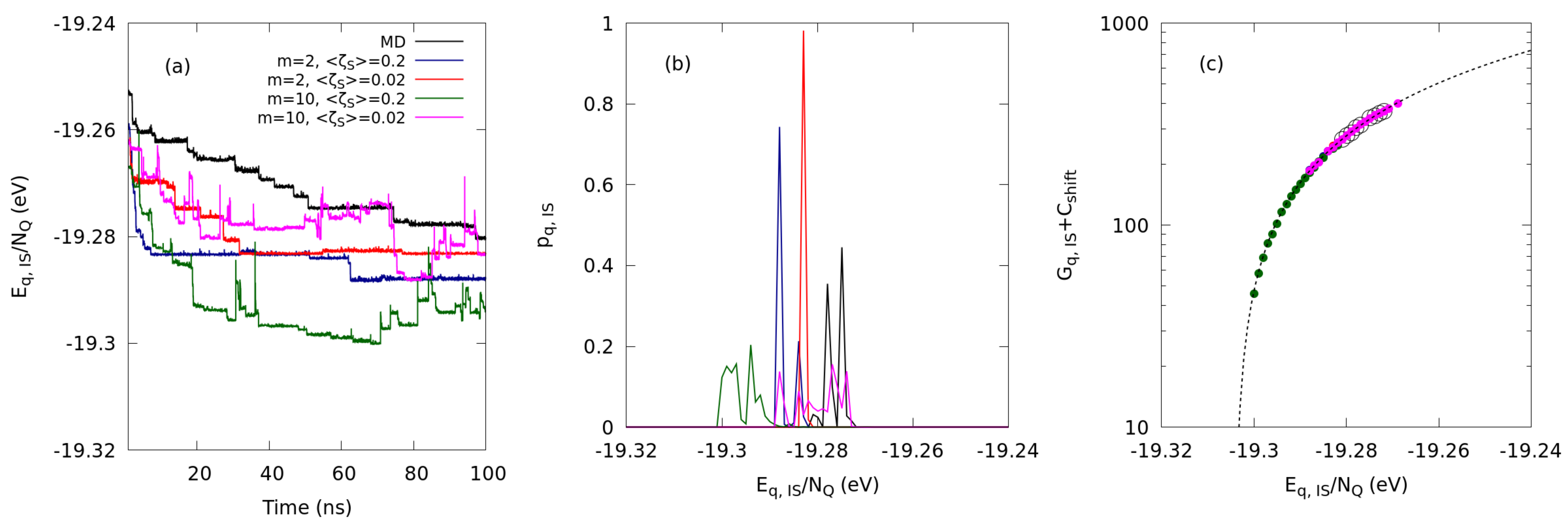}
\caption{Evolution of (a) IS energies of BKS-999 system at different $m$ and $\langle\zeta_S\rangle$ values with time after the initial equilibration phase. (b) $p_{q. IS}$ distribution obtained from the last 50 ns trajectory. (c) $G_{q, IS}$ distribution for different systems. $G_{q, IS}$ distribution for traditional MD system (large empty black circles) is fitted with a Gaussian curve, $G_{q, IS} \propto \exp(-(E_{q, IS}/N_Q - E^0_{q, IS}/N_Q)^2/2\sigma^2)$, to show that all systems faithfully sample the true DOS, with $E^0_{q, IS}/N_Q = -13.5275$ eV, $\sigma = 0.023$ eV. A constant shift factor ($C_{\text{shift}}$) is added to the DOS curves for representational purposes. All energies are represented in per particle scale.}
\label{fig:energy}
\end{figure*}

I have built the EBMD code in C++ language by using LAMMPS~\cite{Thompson_Plimpton_CompPhysComm_2022} software library. I have used Silica as a test system to highlight the potential of entropy-bath accelerated molecular dynamics. I have used the traditional van Beest-Kramer-Santen (BKS)~\cite{Beest_Santen_PhysRevLett_1990} force-field with cutoff modifications as stated in reference~\cite{Saksaengwijit_Heuer_PhysRevLett_2004}. A total of 333 Si particles and 666 O particles are randomly packed in a cubic box with a density of 2.30 g/cc. The system (akin to $Q$-system in the previous section) was equilibrated at a temperature 1000 K using a Nose-Hoover thermostat~\cite{Martyna_Klein_JChemPhys_1994} with 1 fs timestep using LAMMPS~\cite{Thompson_Plimpton_CompPhysComm_2022} software under NVT condition for 1 ns using traditional MD simulation. It is well known that, a glassy material like Silica will show very little dynamics at 1000 K within reasonable timescales of traditional MD simulation, as this temperature is well below its glass transition temperature ($T_g$ = 1475 K)~\cite{Bruning_JNonCrysSolids_2003}. Thus, this system can be taken as a representative of kinetically arrested systems in PEL. 

After the initial equilibration with the thermal bath, I couple the system with the entropy-bath (akin to $S$-system in the previous section). Before starting production runs, one has to make sure that the starting probability distribution maintains $(p_q(t=0)/\Pi_q(t=0)) < 1$ for all values of $\epsilon_q$. If all starting data are taken from a traditional MD simulation, such criteria may be violated. Therefore, a second equilibration scheme is added, where $\langle\zeta_S\rangle$ is slowly increased from zero to the target value for 1 ns. After the two-stage equilibration and proper $p_q(E_q)$ extraction, I simulate the system for production runs with different $m$ and $\langle \zeta_S\rangle$ values for 100 ns and the trajectory is recorded at 100 ps interval. A few parameters are chosen using a trial-and-error method to stabilize the system dynamics. Among the entropy-bath parameters, the $\Delta_s$ factor is assumed to be constant at all $s$ states for simplification. The bath-resetting time $\tau$ was kept at 100 steps. The $M_Q$ factor plays an important role in this study, as it controls the fluctuation of the $p_q$ function and thereby controls the fluctuation of $\vec{F}_q$. A larger $M_Q$ value will lead to a weaker coupling whereas a small $M_Q$ may cause a large fluctuation in $\vec{F}_q$ and break the simulation. Thus, the value of $M_Q$ is reset to $0.5 \times 10^4$ once it reaches a maximum value of $1.0 \times 10^4$ , which seems to work with the current parameter set. In addition, a constant fudge factor ($f_{\zeta}$) is used while updating $\langle\zeta_S\rangle$ as $\langle\zeta_S\rangle = \tilde{\langle\zeta_S\rangle} + f_{\zeta}$ to avoid large fluctuation in $\langle\zeta_S\rangle$ during simulation, with $f_{\zeta} = 5\%\langle\zeta_S\rangle$. The dynamics of the system will also depend on the binning width of the PEL as well since it directly influences the values of $p_q$. A very discrete PEL might lead to undesirable jumps in the energy of the system during the course of the simulation. I have chosen a binwidth of 1.0 eV, which is small enough to create a continuous PEL. Finally, $\pi_q$ values are assumed to be constant at all $q$ states for simplification and estimated as, $\pi_q = \max(p_{q, MD})[m\langle\zeta_S\rangle/(1-\exp(-\beta E_{\text{th}})]^{[1/(m-1)]}$, where $\beta E_{th}=0.05$ is related to an initial estimate of the maximum bias that can be applied per particle. For large $m$ and $\langle\zeta_S\rangle$, the $\vec{F}_q$ fluctuation is often quite large at times, which causes instability in the system dynamics. To account for this effect, the entropy-bath is not allowed to couple with the system if $\beta\Delta V_q > 0.1$ in the successive steps. Therefore, this version of the EBMD algorithm doesn't offer a fixed coupling frequency like a thermal-bath. As proved later, this choice doesn't cause any deviation in estimating $G_q(E_q)$. The total angular and linear momentum is reset to zero at every 10 steps. 

In Fig.~\ref{fig:energy}(a), I show that due to coupling with the entropy-bath, the system escapes the kinetic trap at a much shorter timescale as compared to the traditional MD simulation at the same temperature. Due to a stronger coupling to the entropy-bath, larger $m$ and $\langle\zeta_S\rangle$ values were more efficient in increasing the sampling range of the PEL, as shown in the wider $p_{q,  IS}$ distribution in Fig.~\ref{fig:energy}(b) for $m=10$ systems. The entropy-bath coupling steps for $m=2$ is $100 \%$ in all cases, whereas for $m=10$ systems, it is $\sim 85 \%$ and $\sim 98 \%$ for $\langle\zeta_S\rangle=0.2,0.02$ systems, respectively. The $G_{q, IS}$ distribution was estimated via reweighting the $p_{q,  IS}$ distribution using Eq.~\ref{eqn:Q_distribution_final} and Eq.~\ref{eqn:activation_energy}. It is clear that with $m=10$, the sampling range increased significantly. It can also be seen in Fig.~\ref{fig:energy}(c) that all systems exactly follow that $G_{q, IS}$ distribution predicted by the Gaussian approximation from the traditional MD system. Therefore, the EBMD algorithm described in this latter faithfully samples the true DOS distribution of BKS-999 silica and helps us to explore the regions of PEL that are not accessible in traditional MD simulation at temperatures much lower than $T_g$. 


\section{Conclusion and Outlook}
\label{sect:conclusion}

In this letter, I proved that self-correlation is not necessary to show quantum-like behavior in a closed binary quasi-classical system following a Haldane-type statistical inter-correlation as Eq.~\ref{eqn:nonlinear_haldane_coupled} in between them. This idea is extended to conceptualize an entropy-bath for applications in enhanced sampling techniques in MD simulation. The EBMD algorithm is designed to force a classical system to mimic a quantum-like system, where the maximum occupancy rule via Eq.~\ref{eqn:upper_bound} is enforced. The excellent performance of $m = 10$ systems over traditional MD simulations shows that EBMD simulation can help a kinetically trapped system to escape the high energy barriers and explore low energy basins while maintaining equilibrium even at very low temperatures. 

An important question is: How to choose the proper parameter set for EBMD simulation? Although chosen as constants in this letter, $\pi_q$ and $\Delta_s$ can be a function of energy and may contain multiple adjustable parameters depending on their respective system energies as per the user's choice. The bath-resetting time $\tau$, probability weight $M_Q$, and binwidth of the PEL also play an important role in determining the stability and accuracy of the EBMD simulation. A systematic exploration is underway to optimize the EBMD parameters for various systems, especially biological molecules.  


\section*{Acknowledgement}
The author thanks the National Institute of Technology, Tiruchirappalli, for providing the necessary funding for this research. Necessary simulations were performed in the PARAM Porul supercomputer at NIT-Trichy. I also acknowledge the useful discussions with Prof. Sanjib Senapati and the members of the Computational Bio-Physics Laboratory, Department of Biotechnology, Indian Institute of Technology Madras, India.


\bibliographystyle{apsrev4-2} 
\bibliography{CFES}

\end{document}